# Investigation of Initial Lithiation of Silicon (100) Using Solid-State $^7$Li NMR

Myeonghun Song,[a] Siva P. V. Nadimpalli,[b] Vijay A. Sethuraman,[b] Michael J. Chon,[b] Pradeep R. Guduru,[b] and Li-Qiong Wang[a,*]

[a]Department of Chemistry, Brown University, Providence, Rhode Island 02912, USA
[b]School of Engineering, Brown University, Providence, Rhode Island 02912, USA

*Corresponding Author: Li_Qiong_Wang@brown.edu

**Abstract:** Structural changes in amorphous $Li_xSi$ during the initial lithiation of single crystal Si to amorphous $Li_xSi$ were investigated as a function of Li flux and total charge using solid-state $^7$Li nuclear magnetic resonance (NMR) spectroscopy. Single crystal (100) Si wafers were used as a model system in this study, as the Li flux could be precisely controlled and uniformly distributed across the surface of the wafer. It was observed that peak shifts in solid-state $^7$Li NMR spectra varied as a function of the Li flux during the initial lithiation and stayed constant for samples of the same Li flux regardless of the total charge applied to the electrodes. We conclude from these results that the Li concentration in LixSi stays constant for a given flux regardless of the total coulombic charge applied to the electrode. The results from this study help better understand the kinetics, the reaction mechanisms, and the kinetic modeling of chemical reactions at the reaction front during the initial lithiation of Si (100).

**Introduction**

Silicon, Si, is a promising anode material for use in lithium-ion secondary batteries thanks to its large charge capacity[1], natural abundance and the availability of state-of-the-air fabrication methods. However, performance degradation stemming from mechanical damage driven by colossal volumetric expansion[1, 2] has prevented practical uses of Si-based anodes, though many methods for the improvement of Si anode performances[3-10] have been proposed. Numerous experimental and theoretical studies have provided extensive understanding of the lithiation and delithiation process of crystalline Si to amorphous $Li_xSi$ (where x is the ratio of Li to Si atoms), including phase transformation,[11-15] kinetics,[16-22] mechanical stress and deformation,[23,24] etc. First principle calculations[11,12] predicted that a very small number (x = 0.2 ~ 0.3) of Li atoms could drive the crystalline-to-amorphous phase transition. An *in-situ* X-ray diffraction (XRD) study[15] showed that an amorphous phase of $Li_xSi$ with a fixed Li concentration of x ~ 3.5 is formed during the initial lithiation of crystalline silicon particles. The fixed Li concentration of x ~ 3.5 has been widely accepted in many published studies.[15-17, 24-26] Jung, *et al*.[17] also predicted energetically favorable Li concentration of $Li_{3.4}Si$ at the amorphous-crystalline interface by employing first-principles molecular dynamics simulations. Chon, *et al*.[24] directly observed morphologies of crystalline and amorphous phases and their sharp (~ 1 nm) phase boundary in an electrochemically lithiated single crystal Si wafer by transmission electron microscopy (TEM). Phase transformation, co-existence of crystalline and amorphous phases and a moving phase boundary towards the crystalline silicon phase were considered as essential elements to describe the kinetics[16] for the initial lithiation of crystalline silicon assuming a constant Li concentration of a fixed structure of $Li_{3.5}Si$.

The kinetics of the initial lithiation of crystalline silicon comprise of three processes in series: (i) the redox reaction at the electrolyte/lithiated silicon interface; (ii) the diffusion of lithium through the lithiated phase; (iii) the chemical reaction at the interface between lithiated and crystalline silicon, known as the reaction front. It was found that the chemical reaction at the reaction front is a rate limiting process for the initial lithiation of crystalline silicon,[16] but there is a lack of experimental data for the modeling of kinetics at the reaction front. Therefore, the information on structural changes in $Li_xSi$ as a function of Li flux and total charge is essential for a better understanding of kinetics and reaction mechanism at the reaction front.

Nuclear magnetic resonance (NMR) spectroscopy is a powerful technique for probing local structures of interest, even in a disordered system because it is extremely sensitive to the local magnetic field around nuclei at an atomic scale. Previous *in-situ* and *ex-situ* solid state $^7$Li NMR studies[26-31] on micro- and nano-sized crystalline Si anodes provided information on the local structure of both crystalline and amorphous lithium silicides. Key et al.[26] determined local structures of amorphous $Li_xSi$ formed during the initial lithiation of micro crystalline Si particles using distinct characteristic NMR chemical shifts of known crystalline lithium silicides as references. In both previous NMR studies,[26,27] multiple nuclear magnetic resonances corresponding to different amorphous $Li_xSi$ structures were observed at a given lithiation condition.

Since micro and nano-crystalline Si electrodes were used in previous NMR[26, 27] and XRD[15] studies, the current density (same as Li flux in $\mu A/cm^2$) is not uniform on all particles due to varying particle sizes, shapes and distances from particles to the current collector. In addition, the area of phase boundary decreases as it proceeds into the particle core. Furthermore, a poly-crystalline particle has multiple grains with different crystal orientations and grain boundaries. Hence, it is important to choose a system with a well-defined surface to study the effect of Li flux on the initial lithiation of crystalline silicon.

In this report, structural changes of amorphous $Li_xSi$ were investigated as a function of Li flux and total charge during the initial lithiation of single crystal (100) Si wafers using solid-state $^7Li$ nuclear magnetic resonance (NMR) spectroscopy. Single crystal (100) Si wafers were used as a model system in this study, as the Li flux could be precisely controlled and uniformly distributed across the surface of the wafer. Complementary to the previous NMR studies[26-31] on crystalline Si particles, our objective in this study on single crystal Si wafers is to examine the influence of Li flux and total charge on the structural changes in amorphous $Li_xSi$ during the initial lithiation of crystalline Si. These results will help better understand the kinetics, the reaction mechanisms, and the kinetic modeling of chemical reactions at the reaction front.

**Experimental**

Si electrodes were prepared by cleaving p-type (boron doped) single crystal Si (100) wafer of 250 μm thickness into ~ 1cm x 1 cm squares. A 100 nm Cu film was deposited on the back side of the wafer to form a current collector that will provide uniform current density and prevent lithiation from that side. 1 M LiPF6 in 1:1:1 ethylene carbonate (EC)/diethyl carbonate (DEC)/dimethyl carbonate (DMC) was used as an electrolyte solution and a Celgard polymer film was used as a separator. Coin cells were assembled in an Ar filled glove box with Si as working and Li foil as counter/reference electrodes. The cells were cycled using a Solartron 1470E potentiostat. During lithiation, a Si will react with the Li ions to create a LixSi phase that will uniformly propagate from the front (the side facing the Li) into the bulk Si wafer with a thickness proportional to the total charge of the sample.

To investigate the effect of the Li flux (defined as a current density in $μA/cm^2$) and total charge on the structural changes, the Si electrodes were lithiated at constant current densities of 25 $μA/cm^2$, 50 $μA/cm^2$, 100 $μA/cm^2$ and 200 $μA/cm^2$ for 24 hours and 48 hours (denoted by 25μA24h to 200μA48h). Different lithiation hours gave different total charges at a given current density (*e.g.* 25μA24h *vs*. 25μA48h) and twice the current density at half the lithiation hour had the same total charges at different current densities (e.g. 25μA48h *vs*. 50μA24h).

After finishing the initial lithiation of single crystal Si, coin cells were immediately disassembled and the lithiated Si squares were washed with DMC in the glove box. The washed lithiated Si squares were then cut into 3-4 mm by 6-10 mm pieces and transferred into a 7 mm air-tight NMR sample rotor for angle resolved static NMR measurements. Both angle resolved static NMR and magic angle spinning (MAS) spectra were taken for each lithiated Si wafer using the -air-tight NMR sample rotors (see supporting material for additional experimental details). For MAS NMR measurements, the lithiated Si pieces were cut into smaller pieces (1- 3 mm by 1- 3 mm) and horizontally stacked in a rotor together with inert $SiO_2$ particles to help the magic angle spinning (Fig. 1). A ground sample was made by grinding a lithiated silicon wafer in the glove box using a porcelain mortar.

Solid-state $^7Li$ NMR measurements were carried out with a Bruker DSX 300 spectrometer (7T magnetic field and 116.64 MHz resonance frequency) using both angle resolved and magic angle spinning double resonance probes. For static NMR measurements, the angle between the external magnetic field and the normal direction of the $Li_xSi$ layer ($\theta_n$) was varied by a goniometer. Thus, $\theta_n=0°$ and $\theta_n=90°$ when the external magnetic field is parallel and perpendicular to the normal direction of the $Li_xSi$ layer, respectively. In MAS NMR measurements, the normal direction of the $Li_xSi$ layer in the stacked wafer pieces were parallel to the spinning axis (Fig. 1), therefore the angle of 54.7° between the field and the normal direction was kept constant during spinning at 6.8 kHz. Single pulse NMR spectra were collected with a pulse length of 4 μs (90° pulse) for $^7$Li and a repetition delay of 0.2 s. The number of transients was 1024 ~ 64000 times, depending on the total charge applied to the wafers. $^7$Li NMR chemical shifts were referenced by 1 M LiCl solution in D$_2$O at 0 ppm.

**Results and Discussion**

Fig. 2 shows a constant potential plateaus for the initial lithiation of a single crystal Si at various current densities indicating a reaction limited kinetics with a moving phase boundary, which is consistent with previous studies.[16] It is believed that these potentials are dependent on the Li concentration at the electrolyte/electrode interface and the impedance of a specific cell.. It is observed that the value of the potential plateau decreases with the increasing current density, though there may be up to 20 mV variation for a given current density. This variation may originate from the change in the cell impedance during a long lithiation period. It should be noted that the potential plateau is observed to be very flat in 1-3 hour windows, as is reported in previous studies[16] (see supporting information, S1).

The static NMR spectra (Fig 3) shows the chemical shift for the NMR resonance taken at $\theta_n = 0°$ is higher than that at $\theta_n = 90°$. The angular dependence of the chemical shift, $\delta(\theta_n)$, is due to a demagnetizing field produced inside a magnetic material that cancels out an internal magnetization, known as the bulk magnetic susceptibility (BMS) effect.[32] We observed that $\delta(\theta_n)$ was not affected by the orientation of the underlying crystalline Si wafer, but only to the orientation between the Li$_x$Si layer and the external magnetic field (see supporting information, S2).

The total field experienced by a nuclear spin in the sample can be written a
$$H_{tot} = H_{ext} + H_{int} + M + H_{demag},$$
where $H_{ext}$ is the external field, $M$ is the magnetization, $H_{demag}$ is the demagnetizing field, and $H_{int}$ is the other internal fields except for the magnetization and demagnetizing field. The demagnetizing field is strongly dependent on a sample shape, size and geometrical configuration between the external field and sample. Since the thickness of the Li$_x$Si layer is expected to be smaller than 50 μm in our samples (estimated from the maximum total charge applied), the Li$_x$Si layer can be considered as an infinite plane. It is well known in the classical electrodynamics[33] that $H_{demag} = -M$ when the normal direction of the infinite plane (i.e., the Li$_x$Si layer) is parallel to the external fields and $H_{demag} = 0$ when it is perpendicular to the field. Therefore, the total fields can be described as
$$H_{tot} = H_{ext} + H_{int} \quad (\text{at } \theta_n = 0°)$$
$$H_{tot} = H_{ext} + H_{int} + M \quad (\text{at } \theta_n = 90°).$$

Since the chemical shifts obtained from our study were found to be higher at $\theta_n = 0°$ than at $\theta_n = 90°$, the magnetization $M$ is anti-parallel to the external field, indicating that Li$_x$Si has diamagnetism. From $M = \chi H_{ext}$ and the chemical shift difference of 2 to 4 ppm (see supporting information, S3) between $\delta(\theta_n = 0°)$ and $\delta(\theta_n = 90°)$, the diamagnetic susceptibility $\chi$ is expected to be -2 x 10$^{-6}$ to -4 x 10$^{-6}$. To the best of our knowledge, this is the first experimental observation of magnetic property on the amorphous Li$_x$Si phases. Similar bulk magnetic susceptibility behavior was observed in a paramagnetic Li foil but with an opposite NMR shift direction.[34] In previous NMR measurements [26] on crystalline lithium silicides, the sample shape can be considered a symmetric sphere where the demagnetizing field is $H_{demag} = -1/3M$ [35] regardless of the external field direction. Therefore, the value $\delta_M = \delta(90°) + \frac{1}{3}\{\delta(0°) - \delta(90°)\}$ can be taken as the chemical shifts for the purpose of comparison with chemical shifts of crystalline lithium silicides. NMR chemical shifts plotted as a function of $\theta_n$ are given in Fig. 4 where the chemical shifts were extracted from the peak position in each individual NMR spectrum taken at different angles. Since $\delta(\theta_n)$ resembles the $\cos(2\theta_n)$ function (Fig. 4), $\delta(\theta_n)$ has the value of $\delta_M$ at the magic angle, $\theta_n = 54.7°$.

A few selected spectra from Fig. 3 are plotted in Fig. 5 to better illustrate the chemical shift dependence on Li flux. Figure 5 clearly shows that the chemical shift decreases with the increasing of lithiation current density. NMR spectra given in Figs. 3 and 5 all display a predominant single peak with a linewidth of 1.4 ~ 1.7 kHz (12 ~ 15 ppm) except for 200μA48h, which has a small shoulder at around 18 ppm. The linewidth of the main resonance peak is comparable with that of an individual peak of multiple resonances: ~ 1.0 kHz - 1.4 kHz for micro particles and ~ 1.0 - 2.4 kHz for nano particles in previous MAS NMR studies.[26, 27] A small additional peak at around -3 ppm was observed in Figs. 3 and 5 and its intensity does not depend on the total charge. This peak may be associated with the chemical species in SEI (solid electrolyte interface) that are stuck on the wafer surface and possible second order broadening of quadrupole interactions in static NMR spectra.

In addition to the angle resolved static NMR measurements, MAS NMR measurements at a 6.8 kHz spinning rate were carried out to examine multiple resonance structures which may be buried in our NMR linewidth. At first, samples were ground, but we observed that the grinding process significantly affected the linewidth and peak position, possibly due to the unknown reactions induced by the grinding process (see supporting information, S4). Thus, we adapted an alternative method of horizontally stacking the small $Li_xSi$ wafer pieces in a rotor. In this configuration, the value $\delta_M$ can be obtained directly from the peak position since $\theta_n$ is fixed to the magic angle during spinning. The observed MAS spectra are shown in Fig. 6. It should be noted that the spectra for 25μA24h ~ 100μA24h have an unbalanced base line due to asymmetric spinning side bands (see supporting information, S5). Overall the MAS NMR spectra are similar to those of static NMR spectra. Most spectra except for 200μA48h display a predominant single peak with a linewidth of 1.0 kHz ~ 1.3 kHz and the chemical shift is also systematically decreases with the increasing of current density. However, the shoulder component at around 18 ppm appeared in the static spectrum for 200μA48h is better resolved in the MAS spectrum. This small component observed at 18 ppm is most likely the result of non-uniform lithiation under a high current density of 200 μA/cm.$^2$

Both angle resolved static NMR and magic angle spinning (MAS) spectra for all lithiated Si wafers show a predominant single peak with a relatively narrow linewidth. The line widths of both static and MAS spectra are similar to that of an individual peak from multiple peaks in a spectrum obtained for silicon particles under the high spinning rate of 38 kHz.[26] This suggests that a relatively uniform phase with similar structures of $Li_xSi$ exists in the initial lithiated Si wafer and the MAS NMR spectra taken at 6.5 kHz are adequate to resolve multiple peaks that may be present in this study. Based on the chemical shifts taken for a series of known crystalline $Li_xSi$,[26] the high Li content or higher x in $Li_xSi$ gives a lower chemical shift. This trend is clearly illustrated in the chemical shift curves $\delta(\theta_n)$ as a function of $\theta_n$ and lithiation current density (Fig. 4). The $\delta(\theta_n)$ curves taken at different total charges with the same current density (i.e. 200μA24h and 200μA48h) coincide well with each other, whereas the curves at the same total charge with different current densities are a long way apart. Therefore, it is clear that the Li concentration is largely dependent on the lithiation current density rather than the total charge applied during the initial lithiation of single crystal Si (100) wafer.

In order to check the reproducibility of our measurement and a variation in chemical shift under a given current density, same experiments were repeated several times by making identical coin cells. A similar trend shown in Fig. 4 was found from the results of repeated experiments, and the variation in chemical shifts is not too large to change the overall trend (see supporting information, S3, and the error bar in Fig.7. A few curves that are not matched well in Fig. 4 (e.g. 100μA24h and 100μA 48h) are due to the variation of experimental reproducibility. In comparison with the chemical shifts of crystalline lithium silicides, the $\delta_M$ values taken from both static and MAS spectra are displayed in Fig. 7 as a function of the current density with its variation from the repeated experiments.

According to the previous NMR measurements on crystalline lithium silicides,[26] Li ions associated with a larger number of Si-Si bonds gives a higher chemical shift from 16 ~ 19 ppm (i.e. $Li_7Si_{12}$ and $Li_7Si_3$), while the presence of a larger number of isolated Si ions gives the lower chemical shift, e.g. 6 ppm for $Li_{15}Si_4$ without Si-Si bonds. The chemical shift of 11 ppm for $Li_{13}Si_4$ that contains both isolated Si and Si-Si covalent bonds is located between the lower and higher limit. It is clear that the higher ratio between the number of isolated Si with respect to that of Si-Si bonds corresponds to a lower chemical shift and higher Li concentration in $Li_xSi$. Therefore, the chemical shifts observed in this study are directly correlated with the Si/Si-Si ratio and Li concentrations.

The initial lithiation of crystalline silicon depends on three processes: the redox reaction at the electrolyte/lithiated silicon interface, the diffusion of lithium through the lithiated phase and the chemical reaction at the reaction front, an interface between lithiated and crystalline silicon.[16] Varying $Li_xSi$ structures observed in this study suggests that the value of x in $Li_xSi$ is not fixed at ~ 3.5,[15-17, 24-26] depending on the Li flux. However, the value of x is independent of total charge for a given Li flux. At a higher Li flux, more Si-Si bonds are broken at the reaction front for a given surface area and a given time, giving rise to a larger Si/Si-Si ratio or a higher x in $Li_xSi$. On the other hand, for a given Li flux and higher total charge (or increased lithiation time,), the ratio of Si/Si-Si or the x in $Li_xSi$ remains unchanged due to the moving phase boundary towards the crystalline Si as more Li ions are inserted into the Si electrode.

Complementary to the previous NMR studies[26-31] on crystalline Si particles, this study on single crystal Si wafers enables us to examine the influence of Li flux and total charge on the initial lithiation of crystalline Si due to the simple and well defined geometry of single crystal Si wafer. The previous NMR spectra taken for Si particles[26, 27] show the coexistence of multiple $Li_xSi$ structures during the initial lithiation of Si micro and nano particles at a given lithiation condition. Since micro and nano crystalline Si electrodes were used in previous NMR and XRD[15] studies, the effective current density is not uniform on all particles due to the varying particle sizes and shapes, and distances from particles to the current collector. Therefore, the information on the structural changes in $Li_xSi$ as a function of Li flux and total charge obtained from this study is useful for future modeling of the reaction front kinetics and helps the understanding of the kinetics and mechanisms at the reaction front. However, a thorough description of kinetics at the reaction front requires future studies involving in-situ and time dependent measurements for the initial lithiation of crystalline silicon.

**Conclusions**

In this report, structural changes in amorphous $Li_xSi$ were investigated as a function of Li flux and total charge during the initial lithiation of single crystal (100) Si wafers using solid-state $^7Li$ NMR spectroscopy. A single crystal Si wafer was used as a model system in the study because of well-controlled Li flux uniformly distributed across the surface of a Si (100) wafer. The lithiated Si wafers were produced electrochemically in a coin cell configuration consisting of Si (100) wafer and Li foil as a working electrode and counter/reference electrode, respectively. The initial lithiation of the single crystal Si wafers was achieved galvanostatically at different rates on the coin cells. Both angle resolved static NMR and MAS spectra for each lithiated Si wafer show a predominant single peak with a relatively narrow linewidth, indicating uniform structures of $Li_xSi$. In addition the observed chemical shift systematically decreases with increasing lithiation current density or Li flux, indicating that a higher Li flux results in a higher Li concentration in $Li_xSi$ during the initial lithiation of single crystal Si. It was also observed that the structure of $Li_xSi$ or x in $Li_xSi$ is fixed regardless of the total charge applied at a given Li flux. The results from this study help a better understanding of the kinetics, the reaction mechanisms and the theoretical modeling of the reaction kinetics at the reaction front in the initial lithiation.

**Acknowledgments**

This work was supported by the EPSCoR Implementation Grant DE-SC0007074, Office of Basic Energy Sciences, U. S. Department of Energy (US DOE).

# Figure Captions

**Figure 1:** Sample packing and configuration between sample and external magnetic field for (a) angle resolved static NMR and (b) MAS NMR.

**Figure 2:** Potential profiles during initial lithiation of NMR samples. There is up to 20mV variation in the plateau values which may be attributed to change in cell impedance.

**Figure 3:** Representative angle resolved static NMR spectra taken at $\theta_n = 0°$ and $\theta_n = 90°$.

**Figure 4:** Chemical shift $\delta(\theta_n)$ as a function of lithiation current density and the angle $\theta_n$ between the normal direction of Li$_x$Si layer and the external magnetic field.

**Figure 5:** An array of static NMR spectra taken at $\theta_n = 0°$ and different current densities for (a) 24 h and (b) 48 h. Dashed lines are for the guide of the eye.

**Figure 6:** An array of MAS NMR spectra taken at 6.8 kHz spinning rate and at different current densities and charges.

**Figure 7:** Chemical shift $\delta_M$ as a function of current density observed from (a) static and (b) MAS NMR measurements. Error bar means variation obtained from repeated experiments. Dashed lines correspond to the chemical shift references obtained from crystalline lithium silicides.[26]

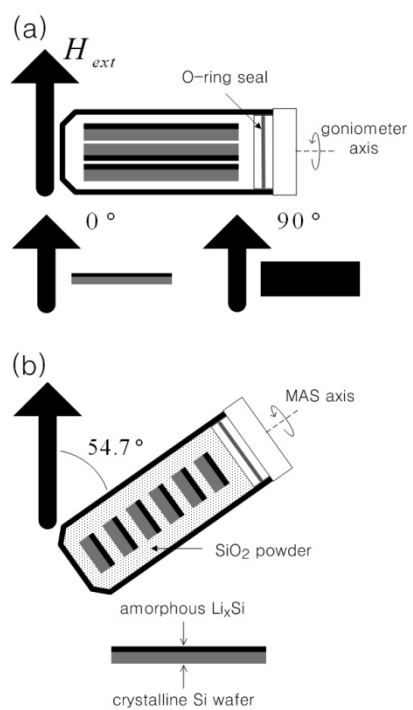

Figure 1: Sample packing and configuration between sample and external magnetic field for (a) angle resolved static NMR and (b) MAS NMR.

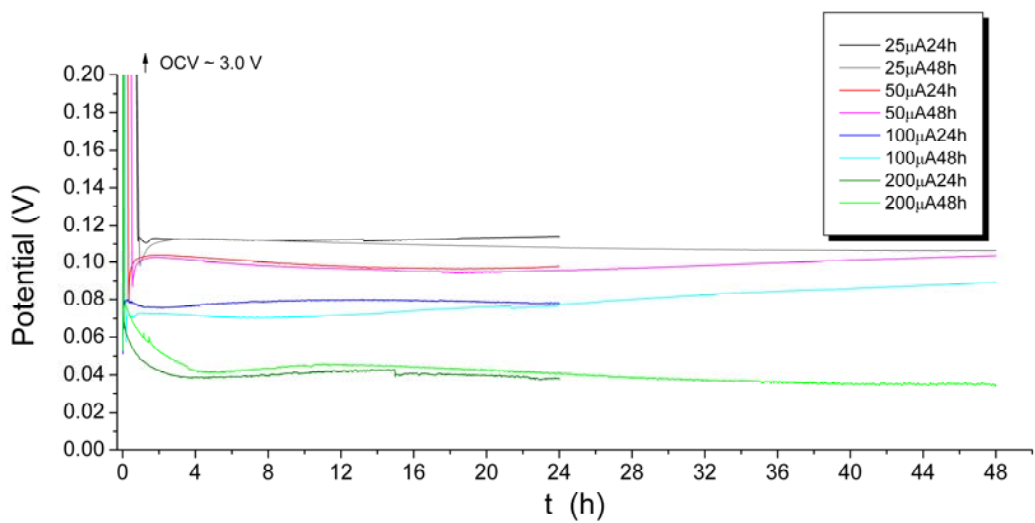

Figure 2: Potential profiles during initial lithiation of NMR samples. There is up to 20mV variation in the plateau values which may be attributed to change in cell impedance.

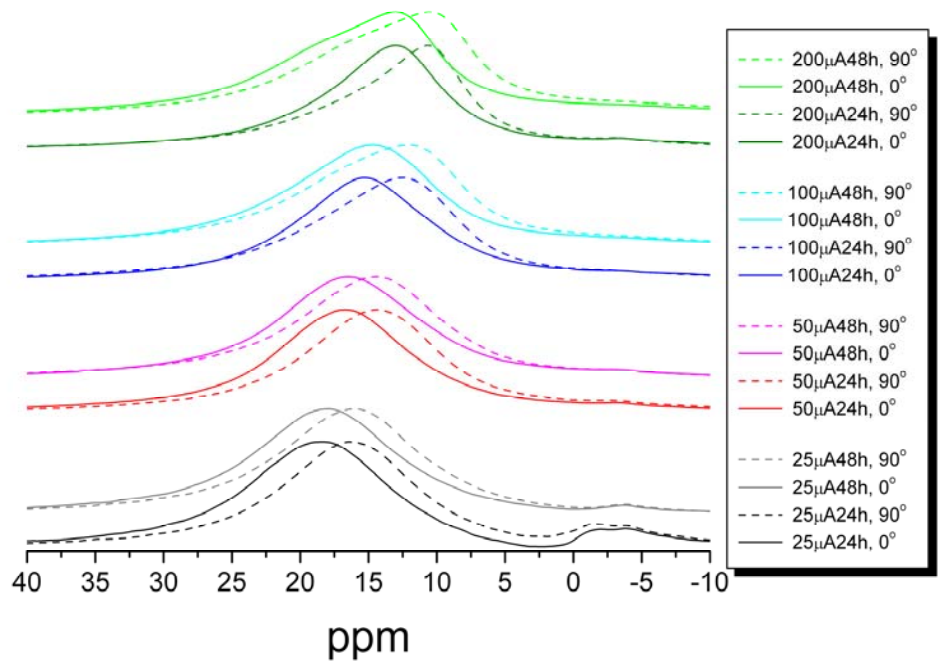

Figure 3: Representative angle resolved static NMR spectra taken at $\theta_n = 0°$ and $\theta_n = 90°$.

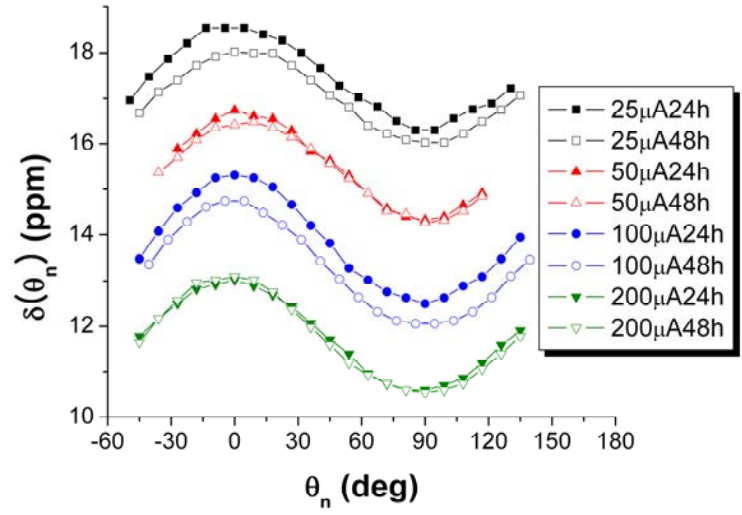

Figure 4: Chemical shift $\delta(\theta_n)$ as a function of lithiation current density and the angle $\theta_n$ between the normal direction of $Li_xSi$ layer and the external magnetic field.

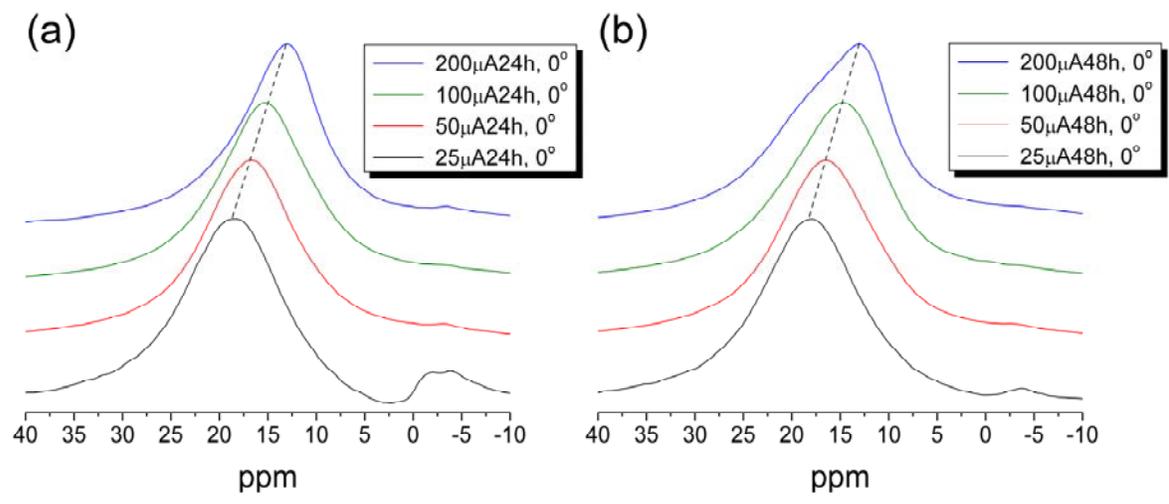

Figure 5: An array of static NMR spectra taken at $\theta_n = 0°$ and different current densities for (a) 24 h and (b) 48 h. Dashed lines are for the guide of the eye.

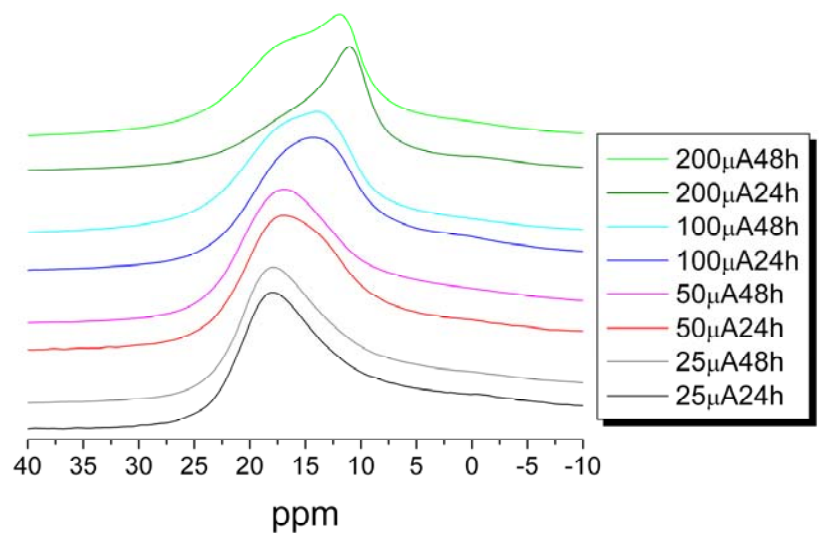

Figure 6: An array of MAS NMR spectra taken at 6.8 kHz spinning rate and at different current densities and charges.

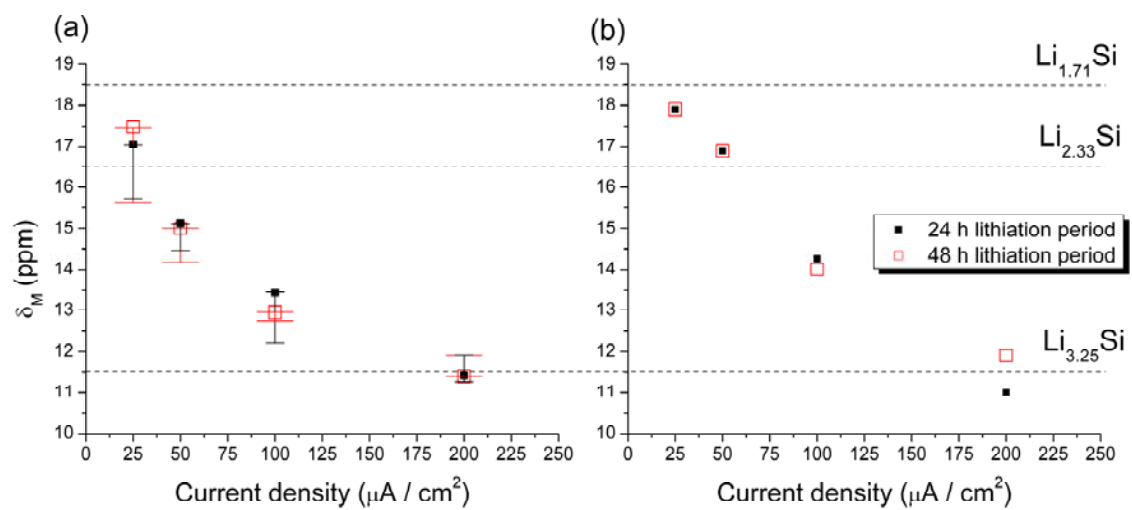

Figure 7: Chemical shift $\delta_M$ as a function of current density observed from (a) static and (b) MAS NMR measurements. Error bar means variation obtained from repeated experiments. Dashed lines correspond to the chemical shift references obtained from crystalline lithium silicides.[26]